\documentclass[superscriptaddress, letterpaper,twocolumn,prl, reprint]{revtex4-1}
\usepackage{amsmath, amsthm, amssymb}
\usepackage[pdftex]{graphicx}
\usepackage{dcolumn} 
\usepackage{bm} 

\setcitestyle{square,super}

\makeatletter 
\gdef\@ptsize{2}
\let\@currsize\normalsize 
\makeatother 
\usepackage{setspace} 
\usepackage[usenames,dvipsnames]{color}

\begin{document}
\title{Atomically precise interfaces from non-stoichiometric deposition}

\author{Y. F. Nie}
\thanks{ These authors contributed equally to this work}
\affiliation{Department of Materials Science and Engineering, Cornell University, Ithaca, New York 14853, USA}
\affiliation{Laboratory of Atomic and Solid State Physics, Department of Physics, Cornell University, Ithaca, New York 14853, USA}
\author{Y. Zhu}
\thanks{ These authors contributed equally to this work}
\affiliation{School of Applied and Engineering Physics, Cornell University, Ithaca, NY 14853, USA}
\author{C.-H. Lee}  
\affiliation{Department of Materials Science and Engineering, Cornell University, Ithaca, New York 14853, USA}
\author{L. F. Kourkoutis} 
\affiliation{School of Applied and Engineering Physics, Cornell University, Ithaca, NY 14853, USA}
\affiliation{Kavli Institute at Cornell for Nanoscale Science, Ithaca, New York 14853, USA}
\author{J. A. Mundy}
\affiliation{School of Applied and Engineering Physics, Cornell University, Ithaca, NY 14853, USA}
\author{J. Junquera}
\affiliation{Departamento de Ciencias de la Tierra y F\'{\i}sica de la Materia Condensada, Universidad de Cantabria, Cantabria Campus Internacional, Avenida de los Castros s/n, 39005 Santander, Spain}
\author{Ph. Ghosez}
\affiliation{Theoretical Materials Physics, Universit\'{e} de Li\`{e}ge, B-4000 Sart-Tilman, Belgium}
\author{X. X. Xi}
\affiliation{Department of Physics, Temple University, Philadelphia, Pennsylvania 19122, USA}
\author{K. M. Shen}
\affiliation{Laboratory of Atomic and Solid State Physics, Department of Physics, Cornell University, Ithaca, New York 14853, USA}
\affiliation{Kavli Institute at Cornell for Nanoscale Science, Ithaca, New York 14853, USA}
\author{D. A. Muller}
\affiliation{School of Applied and Engineering Physics, Cornell University, Ithaca, NY 14853, USA}
\affiliation{Kavli Institute at Cornell for Nanoscale Science, Ithaca, New York 14853, USA}
\author{D. G. Schlom}  
\email[Author to whom correspondence should be addressed: ]{schlom@cornell.edu}
\affiliation{Department of Materials Science and Engineering, Cornell University, Ithaca, New York 14853, USA}
\affiliation{Kavli Institute at Cornell for Nanoscale Science, Ithaca, New York 14853, USA}
\date{\today}

\begin{abstract}

Complex oxide heterostructures display some of the most chemically abrupt, atomically precise interfaces, which is advantageous when constructing new interface phases with emergent properties by juxtaposing incompatible ground states.  One might assume that atomically precise interfaces result from stoichiometric growth, but here we show that the most precise control is obtained for non-stoichiometric growth where differing surface energies can be compensated by surfactant-like effects.  For the precise growth of Sr$_{n+1}$Ti$_n$O$_{3n+1}$ Ruddlesden-Popper (RP) phases, stoichiometric deposition leads to the loss of the first RP rock-salt double layer, but growing with a strontium-rich surface layer restores the bulk stoichiometry and ordering of the subsurface RP structure.  Our results dramatically expand the materials that can be prepared in epitaxial heterostructures with precise interface control---from just the $n=\infty$ end members (perovskites) to the entire RP family---enabling the exploration of novel quantum phenomena at a richer variety of oxide interfaces.

\end{abstract}

\maketitle

 

Members of the Ruddlesden-Popper (RP) homologous series\cite{Balz1955,Ruddlesden:a02086} (see Supplementary Fig. S1) include the high temperature superconductors\cite{ANIEBACK:ANIE198914721},  colossal-magnetoresistance oxides\cite{Moritomo:1996ks},  and systems that with changing dimensionality, controlled by $n$, vary from spin-triplet superconductors\cite{Maeno:1994cm,Mackenzie.75.657} to unconventional ferromagnets\cite{Callaghan1966,shai2013}. Just as the ability to precisely engineer interfaces between perovskites (the $n=\infty$ end member of the RP series) has led to emergent properties at perovskite interfaces, including metal-insulator transitions\cite{Thiel:2006eo,Cen:2008gr}, high electron mobility\cite{Huijben:ADFM201203355} large negative magnetoresistance\cite{Brinkman:2007fk}, and superconductivity\cite{Reyren31082007,Kozuka:2009ht} --- the dimensionality reduction, point defect accommodation\cite{Tilley:1977gz}, and additional novel properties offered by RP phases provide exciting opportunities for oxide interfaces, provided they can be prepared with atomic precision.

Limited by thermodynamic stability, RP materials with $3<n<\infty$ cannot be synthesized by conventional solid state techniques\cite{Balz1955,Ruddlesden:a02086,Udayakumar1988,Udayakumar198955} or codeposition thin film growth techniques\cite{Dijkkamp1987,Koinuma:1987jf,Lathrop:1987hl,Poppe1988661,Sandstrom:1988bm,Spah:1988gi,Madhavan:1996es,Schlom:1999eo,Schlom1998,Matsuno:2003cu,Shibuya:2008if,Tsuyoshi2011}. The ``layer-by-layer" method, in which the species are supplied sequentially following the exact order of monolayers in the growth direction of the targeted structure, has been employed by both pulsed-laser deposition (PLD)\cite{Feenstra:1995ei,Yoshiki1999,Tanaka2000} and reactive molecular-beam epitaxy (MBE)\cite{Schlom100443,Bozovic:ez,LiuZY1994},  to yield well-ordered single-phase epitaxial films of Sr$_{n+1}$Ti$_n$O$_{3n+1}$\cite{Haeni:2001iq, Okude2008},  Sr$_{n+1}$Ru$_n$O$_{3n+1}$\cite{Tian:2007is},  and Ca$_{n+1}$Mn$_n$O$_{3n+1}$\cite{Yan:2007jl}  RP phases with $n$ as high as 10 (ref. 40).  This ability recently led to a breakthrough in tunable microwave dielectric materials\cite{Lee:2013eg}.  In addition to being able to achieve RP phases with intermediate values of $n$, this shuttered method involves significantly lower growth temperatures, compared to the codeposition method. Although sequentially supplying monolayer doses of the constituent species does affect growth in a way that yields the desired layering\cite{Bozovic:ez,Haeni:2001iq,Okude2008,Tian:2007is,Yan:2007jl,LeeCH1023,Locquet1994}, the microscopic details of how the supplied composition modulation affects growth, and more specifically the growth mechanism of RP phases, remains a mystery. This frequently limits the precise control of the interfaces and thus the investigation of oxide interfaces based on RP materials.

\begin{figure*}
\begin{center}
\includegraphics{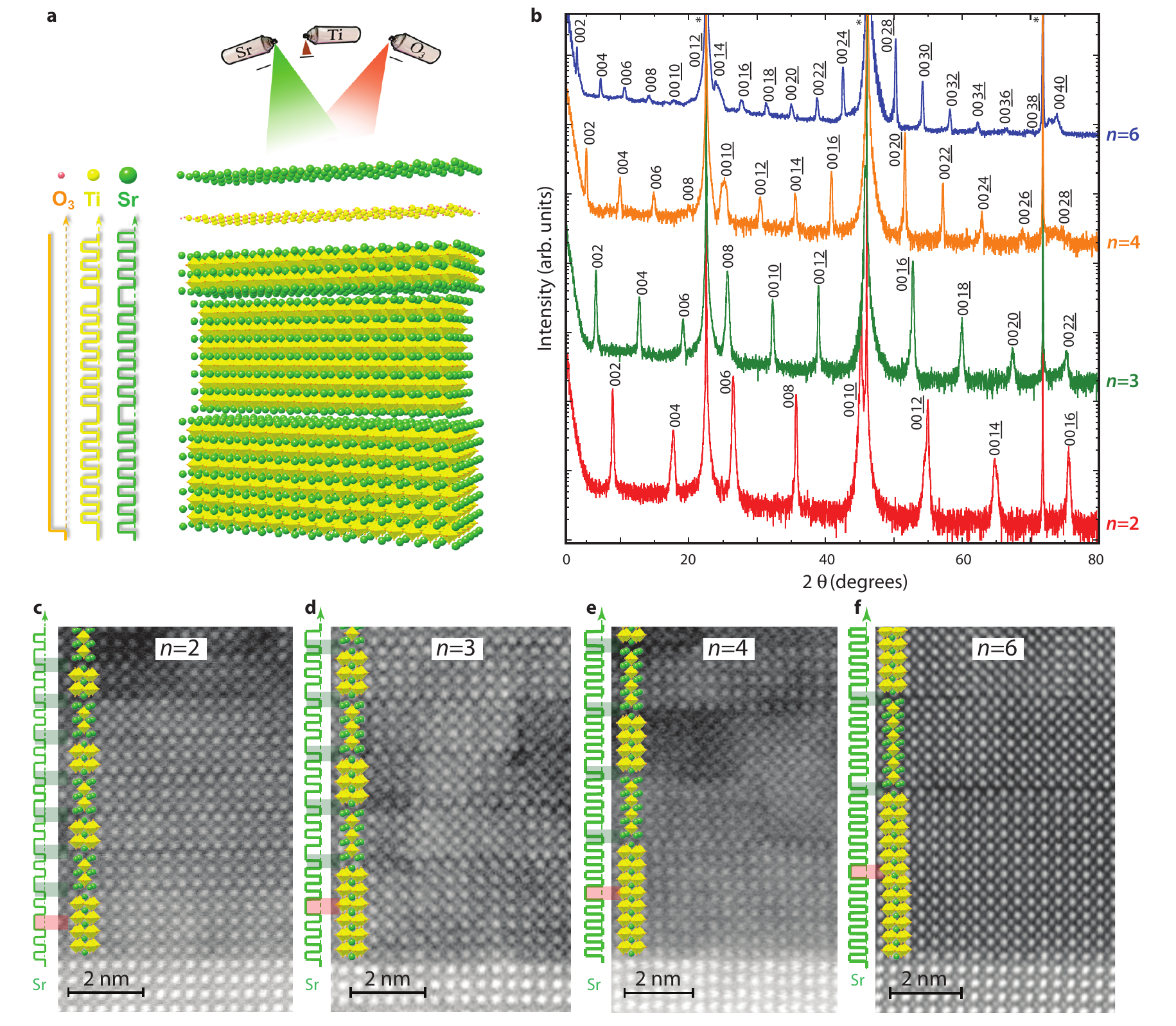}
\renewcommand{\figurename}{Figure}
\caption{\textbf{The missing initial SrO double-layer near the interface of all Ruddlesden-Popper (RP) films.} ({\textbf a}) A spray painting cartoon illustrating the ``layer-by-layer" MBE growth idealization of an $n=6$ RP film. The shutter sequence diagrams for Sr, Ti, and O$_3$ fluxes are shown on the left. ({\textbf b}) $\theta$-2$\theta$ X-ray diffraction of RP films with $n=2,3,4,$ and 6 grown on (110) DyScO$_3$ substrates.  Substrate peaks are labeled with a (*) and the plots are offset for clarity. (\textbf{c-f}) ADF-STEM images from corresponding RP films, showing an unexpected absence of the initial SrO double-layer near the substrate interface of all RP films. The atomic structure model and Sr-flux shutter timing diagram of each film are shown on the left. The positions of present and absent SrO double-layers are highlighted in green and red, respectively. The $n=4$ film shows imperfect growth with some vertical faults and lower contrast between the $A$-site and $B$-site atoms in some regions due to averaging over two grains in the beam direction. The image of $n=6$ RP film was taken on an aberration-corrected Nion UltraSTEM, and thus has higher resolution compared to the other images. 
\label{fig:Missinglayer}}
\end{center}
\end{figure*}


Here, we study the growth mechanism and interface engineering of RP titanate films using a combination of MBE, {\it in situ} reflection high-energy electron diffraction (RHEED), angle-dependent x-ray photoemission spectroscopy (XPS), high-resolution scanning transmission electron microscopy (STEM), and first-principles simulations based on density functional theory (DFT). Interestingly, we observe an unexpected interface between RP films and a substrate with known termination in which the first SrO-SrO rock-salt double-layer (SrO double-layer) is always  missing using the normal ``layer-by-layer" growth process  in which the species are supplied sequentially following the ``layer-by-layer" order of monolayers in the growth direction of the targeted structure. Further, the surface composition, including those designed to be terminated with TiO$_2$, remains SrO. This can be understood as a natural consequence of the formation of a surfactant-like SrO monolayer during the growth of SrO double-layers. We further develop an optimized non-stoichiometric growth process and obtain the desired atomically sharp interface through ``layer-by-layer" engineering. Our work expands the candidate material systems that can be engineered with atomic-layer precision from the $n=\infty$ end member ($AB$O$_3$ perovskites) to the entire RP family, and thus enables the exploration of novel quantum phenomena among a much wider variety of oxide interfaces.

\
\\
\
\textbf{Results}

\noindent\textbf {Missing initial SrO double-layer.} The shuttered ``layer-by-layer" growth method using MBE is schematically shown in Fig.~\ref{fig:Missinglayer}a. During the growth, the fluxes from elemental strontium and titanium sources are shuttered alternately with accurate timing control to provide precise deposition of SrO and TiO$_2$ monolayers onto the (110) DyScO$_3$ substrate in the desired sequence. The oxidant (O$_2$ + ~10\% O$_3$) is supplied during the whole growth and cooling process to minimize the formation of oxygen vacancies. {\it In situ} RHEED is employed to monitor the surface evolution during growth and the intensity oscillations of 	RHEED streaks are used to provide accurate shutter timing calibrations\cite{Haeni2000}. X-ray diffraction (XRD) measurements show intense and narrow peaks and clear thickness (Kiessig) fringes at the expected locations of phase-pure Sr$_{n+1}$Ti$_n$O$_{3n+1}$ with $n=2,3,4,$ and 6 (Fig.~\ref{fig:Missinglayer}b).  The atomic-resolution STEM images confirm the structures consist of $n$ perovskite layers separated by SrO double-layers.  Surprisingly, all samples are missing the initial SrO double-layer (Figs.~\ref{fig:Missinglayer}c-f), showing that the simple ``layer-by-layer" deposition cartoon (Fig.~\ref{fig:Missinglayer}a) is incorrect. Away from the substrate interface, the SrO double-layers are still present, consistent with the XRD results. The missing initial SrO double-layer cannot be explained as simply being due to an imperfect flux calibration. A small deviation (1\%) in the fluxes will lead to clear peak splitting and position shifts in the XRD pattern\cite{Haeni:2001iq, Lee:2013eg,Tian:2007is}, inconsistent with our XRD data. Moreover, the microscopic structure of RP titanates has been shown to be very sensitive to its stoichiometry\cite{Tian:2011bc}. The RP films, other than the $n=4$ one, show very few vertical faults, but all are missing the first SrO double-layer, implying this is not simply because of an imperfect calibration of the monolayer doses supplied to the substrate. Furthermore, it is not related to the surface termination of the substrates. The scandate substrates were prepared to be terminated with ScO$_2$\cite{Kleibeuker:ADFM201000889}. Since the initial deposited layer is SrO, a partial extra SrO layer would be expected if the substrate surface were not perfectly terminated with ScO$_2$, inconsistent with the observation of the  missing SrO double-layer in our case.

\begin{figure}
\begin{center}
\includegraphics{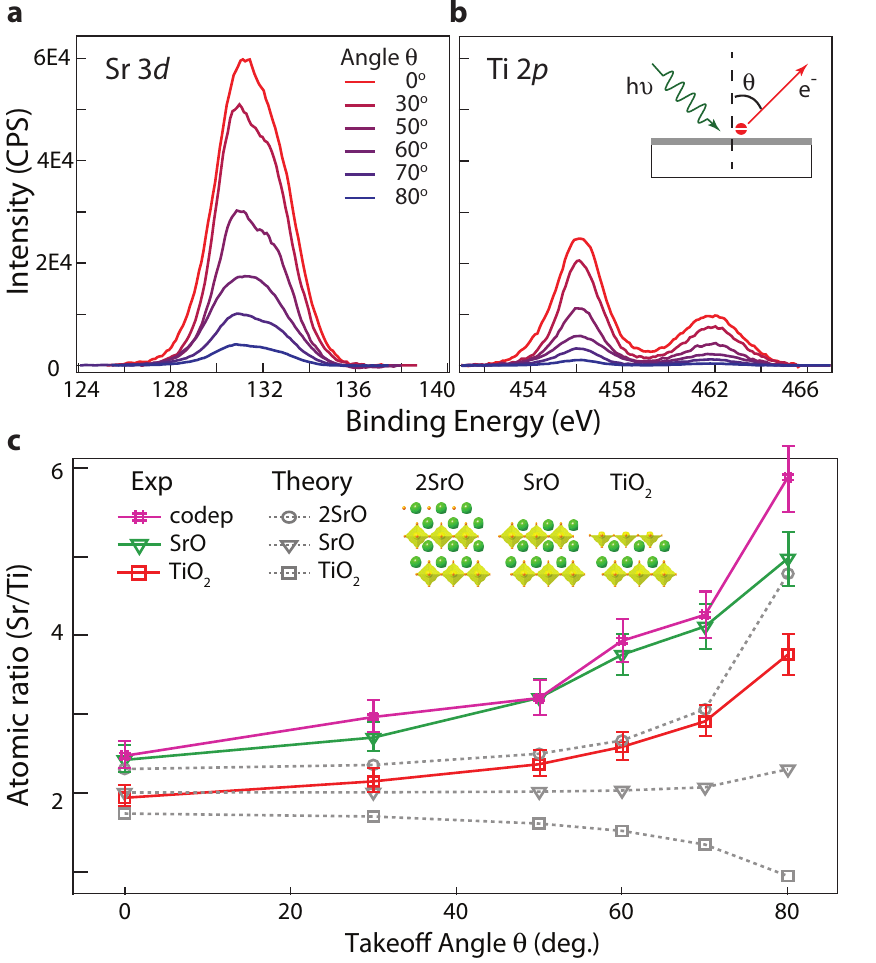}
\renewcommand{\figurename}{Figure}
\caption{\textbf{SrO-terminated surface of RP films  studied by XPS.} Angle-dependent XPS spectra of ({\textbf a}) Sr 3$d$ and ({\textbf b}) Ti 2$p$ peaks of Sr$_2$TiO$_4$ films. Insert: Schematic of the XPS apparatus. ({\textbf c}) Angle-dependent Sr/Ti atomic ratio together with simulations, indicating the surface is terminated with one or more monolayers of SrO for all RP films.
\label{fig:AXPS}}
\end{center}
\end{figure}


\
\\
\
\noindent\textbf {SrO-terminated surface of RP films.} Moreover, we observe an unexpected SrO-terminated surface of all RP titanate films, including those designed to be terminated with TiO$_2$. To study the surface termination, we performed angle-dependent XPS measurements on $n=1$ RP films (Fig.~\ref{fig:AXPS}); the results are representative of RP titanate films containing SrO double-layers, i.e., RPs with finite $n$. Two Sr$_2$TiO$_4$ films were grown on TiO$_2$-terminated  (001) SrTiO$_3$ substrates using the shuttered ``layer-by-layer" growth method and their final deposited layers were designed to be SrO and TiO$_2$, respectively. For comparison, we also grew Sr$_2$TiO$_4$ films using the co-deposition method, where the shutters of all constituent elemental sources were kept open simultaneously during the entire growth process. As the SrTiO$_3$ substrates we used were terminated with TiO$_2$\cite{Koster122630}, the co-deposition method would be expected to yield a TiO$_2$ termination of the film.

\begin{figure*}
\begin{center}
\includegraphics{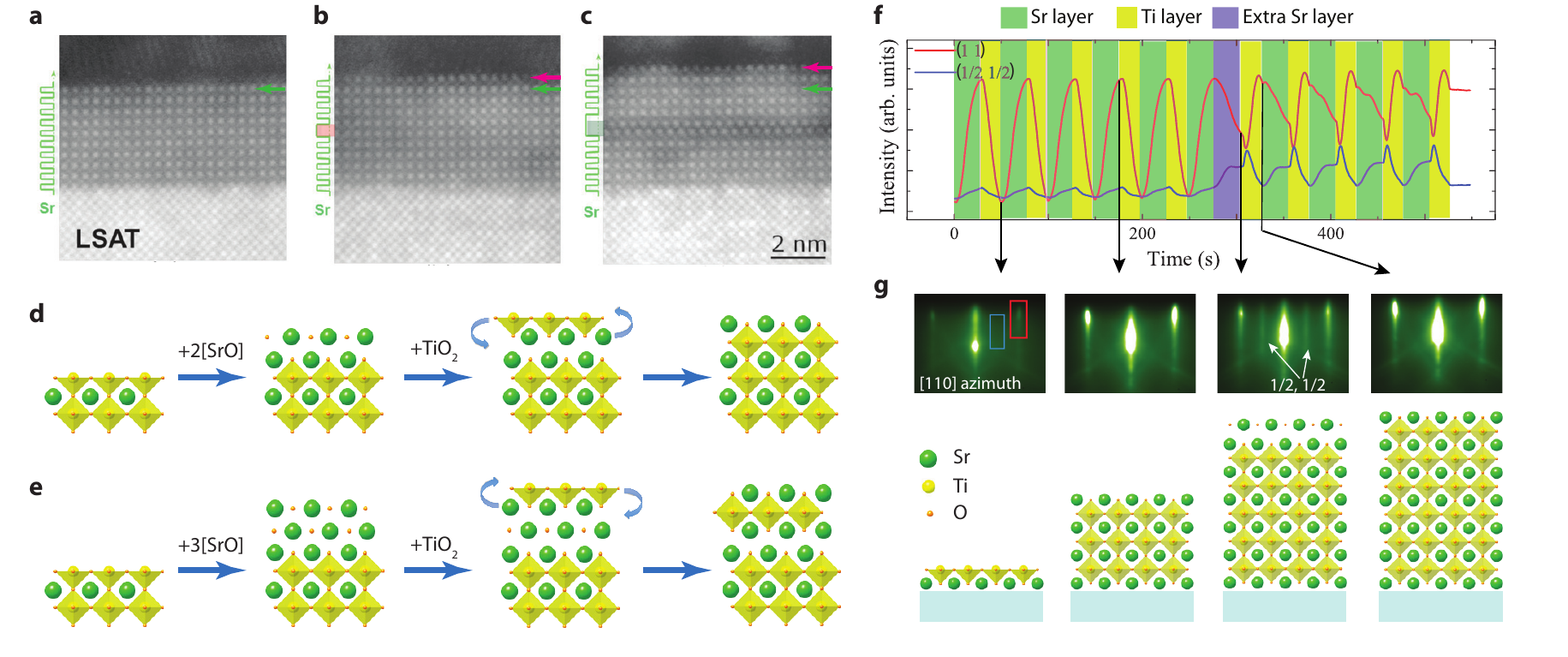}
\renewcommand{\figurename}{Figure}
\caption{\textbf{Microscopic growth mechanism of SrO double-layers.} ADF-STEM images from 10 u.c. thick (001) SrTiO$_3$ films grown on (001) LSAT substrates with an attempt to insert ({\textbf a}) no, ({\textbf b}) 1,  and ({\textbf c}) 2 extra SrO layers halfway through the growth of the 10 u.c. thick SrTiO$_3$ film. A SrO double-layer is formed in the last case only. The excess SrO layer is observed on the film surface, which is indicated by the red arrow. The green arrows indicate the 10th SrO layer. Cartoon models showing the SrO-TiO$_2$ switching mechanism in forming the ({\textbf d}) missing and ({\textbf e}) incorporating SrO double-layer. ({\textbf f}) Shuttered RHEED intensity oscillations of (1, 1) and (1/2, 1/2) diffraction streaks within the red and blue boxes, respectively, during the growth of the artificial structure in ({\textbf b}), showing the evolution of the surface diffraction pattern along the [110] azimuth of SrTiO$_3$. ({\textbf g}) RHEED patterns along the [110] SrTiO$_3$ azimuth and corresponding cartoon models taken at different stages of growth indicated by the arrows. 
\label{fig:growthdynamics}}
\end{center}
\end{figure*}

In Figs.~\ref{fig:AXPS}a,b, the XPS profile of  Sr $3d$ and Ti $2p$ peaks show a clear dependence on the takeoff angle.  The area of each peak is calculated after subtracting a Shirley background\cite{shirley1972high} (not shown) and is used to determine the atomic concentration for that element after being scaled by a relative sensitivity factor. The extracted atomic Sr/Ti ratios clearly increase with takeoff angle $\theta$ for all films (Fig.~\ref{fig:AXPS}c). Comparison to the theoretical simulations (dashed lines) implies that all films are terminated with one or more SrO layers. The clear angular dependence of the atomic Sr/Ti ratios unambiguously implies a SrO-terminated surface for all of the Sr$_2$TiO$_4$ films, even those attempted to be TiO$_2$ terminated. 

\
\\
\
\noindent\textbf {Microscopic growth mechanism of SrO double-layers.} To reconcile these unexpected observations of a missing interfacial SrO double-layer and the ever-present SrO-terminated surface, we investigate the microscopic growth mechanism of RP films, more specifically the SrO double-layer, by studying the growth of a single SrO double-layer in thick SrTiO$_3$ films. To avoid strain effects and the introduction of defects from relaxation,  we used (001) (LaAlO$_3$)$_{0.29}$(SrAl$_{0.5}$Ta$_{0.5}$O$_3$)$_{0.71}$ (LSAT) substrates since they are well lattice matched with RP titanates  with low $n$. After growth, a 10 nm thick amorphous La$_2$O$_3$ capping layer was deposited on the film surface at room temperature to protect the surface for high-resolution STEM measurements. We first grew a 10 unit cell (u.c.) thick (001) SrTiO$_3$ film on (001) LSAT to verify the growth conditions. Indeed, the STEM image shows the	 SrTiO$_3$ to have the expected 10 u.c. thickness by counting the number of SrO layers (Fig.~\ref{fig:growthdynamics}a). For the second sample, we attempted to insert a SrO double-layer into the middle of the 10 u.c. thick SrTiO$_3$ structure by depositing one extra SrO layer after the first 5 u.c. of SrTiO$_3$. Surprisingly, the final structure shows no trace of a SrO double-layer at any location (Fig.~\ref{fig:growthdynamics}b). Instead, we find an extra SrO layer on the top surface of the film, suggesting the inserted extra SrO layer switches position with the following TiO$_2$ layer as schematically shown in the cartoon model (Fig.~\ref{fig:growthdynamics}d). This position switching process continues until the end of the film growth, resulting in an extra SrO monolayer on the top surface. The extra SrO monolayer ``floats" on the surface of the growing film and is not incorporated, analogous to a surfactant.

This hypothesis is further supported by our real-time {\it in situ} RHEED measurements (Figs.~\ref{fig:growthdynamics}f,g). RHEED patterns are very surface sensitive and the maximum and minimum intensity of the (1,1) diffraction streak correspond to the SrO-terminated and TiO$_2$-terminated surfaces of SrTiO$_3$, respectively\cite{Haeni2000}. One period of the observed shuttered intensity oscillation corresponds to the growth of a full u.c. of SrTiO$_3$. In contrast, the presence of half-integer RHEED streaks at positions along the [110] azimuth of (001) SrTiO$_3$ indicate excess SrO (more than one full layer) on the surface\cite{Yu200451, Fisher2008JAP, Lee2013APL}. In Fig.~\ref{fig:growthdynamics}f, the shuttered RHEED intensity oscillation shows a significant increase of the (1/2, 1/2) peak during the deposition of the extra SrO layer, suggesting an increase of excess SrO on the surface of the film. After depositing the following TiO$_2$ layer, however, the intensity of (1/2, 1/2) streaks decreases to a minimum, while the intensity of (1, 1) streaks goes back to a maximum value, identical to the SrO-terminated surface in a perfect SrTiO$_3$ structure. The corresponding RHEED pattern is also identical to the SrO-terminated surface, implying the SrO layer indeed switched position with the TiO$_2$ layer. During the growth of the last 5 u.c. of SrTiO$_3$ of this second sample, the intensity of the (1/2, 1/2) and (1, 1) peaks oscillate periodically in an anti-phase manner, suggesting that such a SrO-TiO$_2$ switching process continues to take place until the end of growth, leaving a SrO-terminated surface. This is consistent with our STEM measurements. 


\begin{figure}
   \begin{center}
      \includegraphics{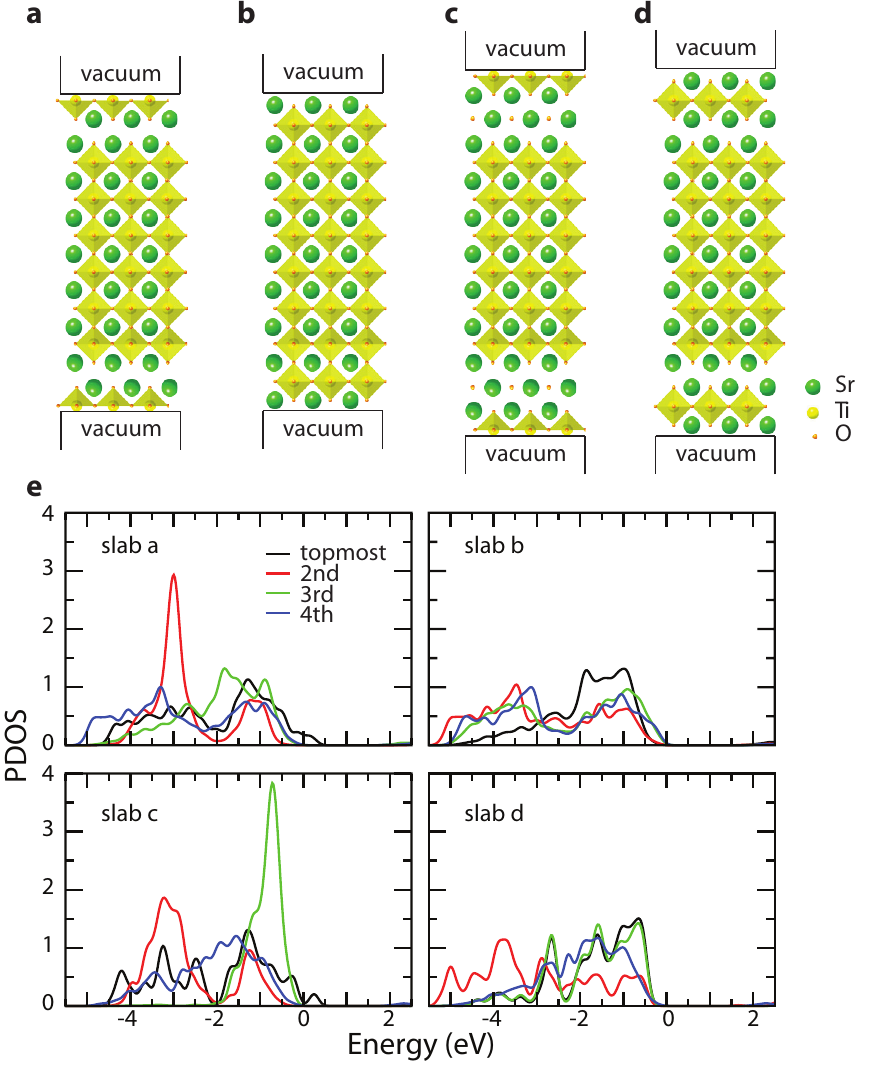}
      \renewcommand{\figurename}{Figure}
      \caption{\textbf{Unfavorable TiO$\boldsymbol{_2}$-terminated surface from DFT calculations.} (\textbf{a-d}) Schematic diagram of the slabs considered, having various terminations.     
                ({\textbf e})  Partial density of states projected onto the O $2p$ 
                orbitals of the topmost four layers from the surface inward.
                The zero energy level is the top of the valence band 
                computed from the position of the local band edge at the 
                central SrO layer in each slab. The Fermi energy always lies
                within the gap.
      \label{fig:DFT}}
   \end{center}
\end{figure}


This behavior can also be understood from energy considerations. In RP structures with low $n$, the distance between SrO and TiO$_2$ layers is 1.92 \AA, while the distance between two SrO layers is 2.56 \AA\cite{Venkateswaran19871273}, which generally would imply a weaker bond energy between two such rock-salt planes. This is a common feature of all RP materials. As a consequence, the natural cleavage plane of RP phases, such as layered manganites\cite{Loviat2007} and ruthenates\cite{Matzdorf:2000jm}, is between the rock-salt double-layers.  Thus, we expect a TiO$_2$ layer deposited on top of an SrO double-layer to find a lower energy configuration by switching positions with the uppermost SrO layer and residing between them where it can bond with both SrO layers. 

Indeed, our DFT calculations support this qualitative notion and show that the structures after switching  of TiO$_2$ and SrO layers are energetically more favorable than before the switching (Fig.~\ref{fig:DFT}).  
 Four different types of free surfaces have been simulated using 
 periodic symmetrically-terminated slabs, schematically shown in 
 Figs.~\ref{fig:DFT}a-d. 
 Two of them (Figs.~\ref{fig:DFT}a,b) contain 9 layers of SrO and 8 layers of 
 TiO$_{2}$, while the other two (Figs.~\ref{fig:DFT}c,d) contain 11 layers of 
 SrO and 8 layers of TiO$_{2}$.
 For each pair of structures with the same overall stoichiometry, the first TiO$_{2}$ layer has been placed
 in different positions, corresponding to the possible diffusion of titanium into the
 underlying structure: 
 (i) at the top most atomic layer with two consecutive
 layers of SrO beneath it (Figs.~\ref{fig:DFT}a,c), and
 (ii) at the subsurface layer, in between two SrO planes (Figs.~\ref{fig:DFT}b,d).
 The vacuum regions were thicker than 12 {\AA} 
 (more than three unit cells of SrTiO$_3$), large enough to avoid interactions 
 between periodic replicas.  
 Symmetric slabs were used in order to only focus on the effects 
 of the surfaces. 
 The in-plane lattice constants were chosen to match the theoretical lattice constant of bulk SrTiO$_{3}$ (3.874 \AA)~\onlinecite{Junquera-03.2}.
 
The calculated total free energy differences ($\Delta E$) are 
 0.76 eV between slab (a) and (b), and 1.19 eV between slab (c) and (d), 
 implying a large energy gain by surrounding TiO$_2$ with SrO layers 
 on both sides.
%
 These energetic considerations support the experimental observation
 about the tendency of the TiO$_{2}$ layers to diffuse down into the 
 structure.
 They can be understood from
 the evolution of the character of the bondings
 with the position of the TiO$_2$ layer in the different
 structures.

 In Figure~\ref{fig:DFT}(e) we plot the layer-by-layer 
 projected density of states (PDOS) on an oxygen atom
 located in one of the four topmost layers.
 Slabs (a) and (c), with TiO$_2$ planes at the interface with vacuum, display states with energies just
 above the top of the bulk valence band with nearly pure
 O 2$p$ character (without hybridization with Ti 3$d$ levels).
 In fact, at the top of the bulk valence band, all of the O 2$p$ orbitals
 are hybridized to some extent with Ti 3$d$ states, which lowers the energy
 of the O 2$p$ orbitals.
 At the TiO$_{2}$-terminated surface, 
some of these bonds are missing; the mixing
 of the states that shift the position of the valence
 bands is no longer possible. Therefore, O 2$p$ states remain close
 to the free-atom energy levels, intruding into the lower part of the
 gap. On the other hand, when the surface is SrO-terminated, every surface oxygen-atom
 has a titanium atom below it with which it hybridizes, resulting in a lowering of
 the PDOS towards higher binding energies.
 In addition, when the topmost TiO$_{2}$ layer is surrounded by two SrO layers,
 as occurs in slabs (b) and (d), the PDOS on the SrO planes
 more closely resembles that of the SrO planes in bulk SrTiO$_3$ than that of bulk SrO.
 This translates into a wider shape of the PDOS,
 spreading over higher binding energies. As a consequence of the partial covalent bonding in the perovskite, the sharp peaks located around 3.0 eV [slab (a)], and 1.0 eV and 3.0 eV [slab(c)] below the top of the valence band,
 which signify the strong ionic bonding in SrO, are distributed
 over larger energy windows in slab (b) and (d), respectively.
 Clearly, the energy of the TiO$_2$ plane will be lowered
 if it is not at the surface. Experimentally, the actual structures may have various types of 
 defects and surface reconstructions. We should also point out that our simulations do not address
 the dynamical problem of the diffusion. Nonetheless, our calculations indicate that a TiO$_2$-terminated 
 surface atop more than one SrO layer is unstable with respect to inward migration and
 agrees qualitatively with our experimental observations.


\begin{figure}
\begin{center}
\includegraphics{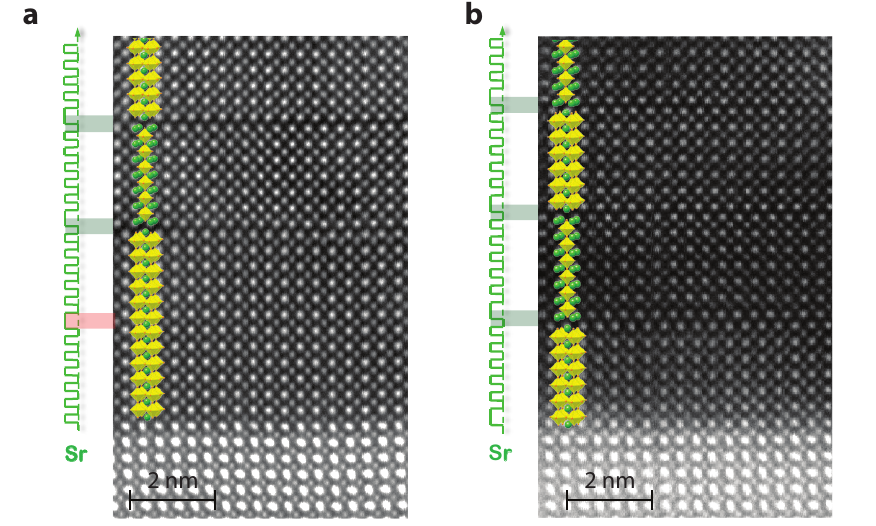}
\renewcommand{\figurename}{Figure}
\caption{\textbf{Achievement of the desired interface in $\boldsymbol{n=6}$ RP films through ``layer-by-layer" engineering.} ADF-STEM images from $\boldsymbol{n=6}$ RP films grown on (110) DyScO$_3$ using (\textbf{a}) conventional and (\textbf{b}) revised shutter sequences, showing the successful growth of the targeted interface in the latter non-stoichiometric case. The positions of present and absent SrO double-layers are highlighted in green and red, respectively.
\label{fig:sharpinterface}}
\end{center}
\end{figure}


\
\\
\
\noindent\textbf {Atomically precise interface control.} In order to obtain the desired control over where the SrO double-layer interfaces occur with atomic specificity, we grew a third sample into which 2 extra SrO layers were inserted into the middle of a 10 u.c. thick SrTiO$_3$ structure. As shown in Fig.~\ref{fig:growthdynamics}c, a SrO double-layer appears at this location and an extra SrO layer is found on the surface. This  suggests that the TiO$_2$ layer deposited subsequently to the extra 2 SrO layers only switched position with the top most SrO layer, leaving a SrO double-layer (Fig.~\ref{fig:growthdynamics}e) as shown in STEM measurements (Fig.~\ref{fig:growthdynamics}c). This is again supported by an energy consideration in our DFT calculations (Fig.~\ref{fig:DFT}). Guided by this cation switching model, we establish an optimized growth recipe for RP titanate films by depositing one extra SrO layer at the substrate interface to form the one-SrO-monolayer-rich condition necessary for the controlled growth of subsequent SrO double-layers at desired positions with atomic-layer specificity. In Fig.~\ref{fig:sharpinterface}, an $n=6$ RP film grown by this optimized non-stoichiometric deposition process indeed shows the desired interface.

\
\\
\
\textbf{Discussion}

Through a series of experiments and calculations we have elucidated the microscopic growth mechanism of SrO double-layers in RP phases, which explains  observations counter to the cartoon ``layer-by-layer" growth model (Figs.~\ref{fig:Missinglayer}a). It has long been assumed that RP phases grow in the same Òlayer-by-layerÓ order that their constituent monolayers are deposited, but our results show this is untrue.  We further show that a desired SrO double-layer can be inserted with atomic layer control into RP titanates through an optimized growth process. This demonstrates the power of non-stoichiometric deposition through ``layer-by-layer" engineering in growing atomically sharp interfaces at desired locations in complex oxide heterostructures. Our work provides a route to study novel interfacial phases containing any desired $n$ member within the RP series. 

\
\\
\
\noindent\textbf{Methods}\\
\noindent\textbf{Reactive molecular-beam epitaxy (MBE) film growth:} Sr$_{n+1}$Ti$_n$O$_{3n+1}$ RP titanate films with $n$ up to 6 were deposited  at a substrate temperature of 750-780~$^\circ$C and an oxidant background pressure (O$_2$ + $\sim$10\% O$_3$) of $3\times10^{-7}$ -  $5\times10^{-7}$ Torr using Veeco 930 and Gen 10 oxide MBE systems.  A low-temperature effusion cell and a Ti-Ball$^{\mathrm {TM}}$ (ref. 55) were used to generate strontium and titanium molecular beams, respectively. A shuttered ``layer-by-layer" method was used to grow all samples in this work. The fluxes of strontium and titanium were initially calibrated using a quartz crystal microbalance and the precise monolayer shutter timing of the strontium and titanium doses were subsequently refined using RHEED intensity oscillations\cite{Haeni2000}.
\\
\
\\
\noindent\textbf{Scanning transmission electron microscopy (STEM):} Samples for STEM measurements were prepared by mechanical wedge polishing followed by short-time Ar-ion milling\cite{Voyles2003251}.  High-angle annular-dark-field (HAADF) STEM images were taken both on a 200 kV FEI Tecnai F20 SuperTWIN and on a 100 kV Nion UltraSTEM equipped with an aberration corrector\cite{Kirkland20111523}. STEM images, except the one shown in Fig. 1e ($n=4$), were averaged over a few frames (pre-aligned using cross-correlation algorithm) to reduce the scan distortion.
\\
\
\\
\noindent\textbf{X-ray photoemission spectroscopy (XPS):} The {\it ex situ} angle-dependent XPS measurements were performed under ultrahigh-vacuum (UHV) using monochromatic Al K$_\alpha$  x-ray radiation. Samples were stored under ultra-high vacuum before XPS measurements to minimize the exposure of the samples to air. To prevent surface charging during the XPS measurements, the samples were irradiated with low energy electrons. In angle-dependent XPS, the photoelectron yield decreases exponentially with sampling depth at a fixed takeoff angle. The total intensity for photoelectrons $I_j(\theta)$, is a weighted sum of the signals emanating from all sampling depths\cite{Livesey1994439}.
\begin{eqnarray}
I_j({\theta})=K_j\sum_{i=0}^{\infty}n_{j,i}\mathrm{exp}({-Z_{i}\over{\lambda}_j\mathrm{cos}{\theta}}),
\label{Eq1}
\end{eqnarray}
where $K_j$ is a constant related to the elemental sensitivity and instrumental factors of atom $j$, $n_{j,i}$ represents the atom fraction composition of element $j$ in the $i$th layer, $Z_{i}$ is the depth of the $i$th layer from the surface, $\lambda_{j}$ is the characteristic attenuation length for the photoelectrons originating from element $j$, and $\theta$ is the takeoff angle of emission measured from the sample normal. The characteristic attenuation length   $\lambda_j$ for the photoelectrons originating from Sr $3d$ and Ti $2p$ are calculated to be 25.69 \AA~ and 20.77 \AA, respectively, using the NIST Electron Effective-Attenuation-Length database\cite{EAL}. The instrument-dependent constant $K_j$ can be cancelled out in calculating the atomic concentrations and ratios\cite{Livesey1994439}.
\\
\
\\
\noindent\textbf{DFT	 calculations:} 
DFT calculations, in the local density approximation (LDA),
 were performed within a numerical atomic orbital method,
 as implemented in the {\sc Siesta} code~\cite{siesta}.
 Core electrons were replaced by fully-separable 
 norm-conserving pseudopotentials.
 Due to the large overlap between the semicore and valence states,
 the $3s$ and $3p$ electrons of titanium, and
 $4s$ and $4p$ electrons of strontium were considered as valence electrons
 and explicitly included in the simulations.
 The one-electron Kohn-Sham eigenvalues were expanded in a basis
 of strictly localized numerical atomic orbitals.
 We used a single-zeta basis set for the semicore states of titanium and strontium,
 and a double-zeta plus polarization for the valence states of all the atoms.
 For strontium, an extra shell of 4$d$ orbitals was added.
 Further details on the pseudopotentials and basis sets
 can be found in ref. 60.
 The electronic density, Hartree and exchange-correlation potentials,
 as well as the corresponding matrix elements between the basis orbitals, were
 computed on a uniform real space grid with an
 equivalent plane-wave cutoff of 1200 Ry in the representation of the
 charge density.
 A $ 12 \times 12 \times 1 $ Monkhorst-Pack 
 mesh was used for the sampling of reciprocal space.
To compute  the PDOS,
 a non-self-consistent calculation within a grid of
 $60 \times 60 \times 2$ was employed.  To optimize the geometry of the structure, atomic positions
 were relaxed until the value of the maximum
 component of the force on any atom fell below 10 meV/\AA.


\textbf{Acknowledgments}

This research was supported by Army Research Office (ARO) grants W911NF-12-1-0437 (for C.-H.L., J.A.M., and D.G.S.) and W911NF-09-1-0415 (for Y.Z. and D.A.M.) and by the National Science Foundation through the MRSEC program (DMR-1120296 for Y.N. and and L.F.K.). Y.Z. and L.F.K. thank J. L. Grazul, M. G. Thomas, and E. J. Kirkland for maintaining the microscope. This work made use of the electron microscopy facility of the Cornell Center for Materials Research (CCMR) with support from the National Science Foundation Materials Research Science and Engineering Centers (MRSEC) program (DMR 1120296) and NSF IMR-0417392. This work was performed in part at the Cornell NanoScale Facility, a member of the National Nanotechnology Infrastructure Network, which is supported by the National Science Foundation (Grant ECCS-0335765). Ph.G. acknowledges a Research Professorship from the Francqui Foundation and financial support of the ARC project ``TheMoTherm.'' J.J. acknowledges finantial support from the Spanish Ministry of Economy and Competitiveness through Grant No. FIS2012-37549-C05-04. J.J. gratefully acknowledges the computer resources, technical expertise and assistance provided by the Red Espa\~nola de Supercomputaci\'on.

\textbf{Author contributions}

Y.N., Y.Z. and D.G.S. designed the experiments. Y.N. and C.-H.L. grew the films and performed the {\it in situ} measurements and structural characterizations. Y.N. performed the XPS measurements and analyzed the data. Y.Z., L.F.K. and J.A.M. performed the STEM measurements and analyzed the data. J.J., Y.N. and Ph.G. performed the first principle calculations and analyzed the data.  D.G.S., D.A.M., K.M.S, X.X.X. and Ph.G. supported and supervised the project. Y.N., Y.Z. and D.G.S. wrote the manuscript. All authors reviewed and edited the manuscript.

\textbf{Additional information}

Supplemental information is available in the online version of the paper. Reprints and permissions information is available online at www.nature.com/reprints. Correspondence and requests for materials should be addressed to D.G.S.

\textbf{Competing Financial Interests}

The authors declare no competing financial interests.


\begin{thebibliography}{10}
\expandafter\ifx\csname url\endcsname\relax
  \def\url#1{\texttt{#1}}\fi
\expandafter\ifx\csname urlprefix\endcsname\relax\def\urlprefix{URL }\fi
\providecommand{\bibinfo}[2]{#2}
\providecommand{\eprint}[2][]{\url{#2}}

\bibitem{Balz1955}
\bibinfo{author}{Balz, D.} \& \bibinfo{author}{Plieth, K.}
\newblock \bibinfo{title}{{Die struktur des Kaliumnickelfluorids,
  K$_2$NiF$_4$}}.
\newblock \emph{\bibinfo{journal}{Z. Elektrochem.}}
  \textbf{\bibinfo{volume}{59}}, \bibinfo{pages}{545--551}
  (\bibinfo{year}{1955}).

\bibitem{Ruddlesden:a02086}
\bibinfo{author}{Ruddlesden, S.~N.} \& \bibinfo{author}{Popper, P.}
\newblock \bibinfo{title}{{New compounds of the K${\sb 2}$NiF${\sb 4}$ type}}.
\newblock \emph{\bibinfo{journal}{Acta Crystallogr.}}
  \textbf{\bibinfo{volume}{10}}, \bibinfo{pages}{538--539}
  (\bibinfo{year}{1957}).

\bibitem{ANIEBACK:ANIE198914721}
\bibinfo{author}{M\"uller-Buschbaum, H.}
\newblock \bibinfo{title}{{The crystal chemistry of high-temperature oxide
  superconductors and materials with related structures}}.
\newblock \emph{\bibinfo{journal}{Angew. Chem. Int. Ed. Engl.}}
  \textbf{\bibinfo{volume}{28}}, \bibinfo{pages}{1472--1493}
  (\bibinfo{year}{1989}).

\bibitem{Moritomo:1996ks}
\bibinfo{author}{Moritomo, Y.}, \bibinfo{author}{Asamitsu, A.},
  \bibinfo{author}{Kuwahara, H.} \& \bibinfo{author}{Tokura, Y.}
\newblock \bibinfo{title}{{Giant magnetoresistance of manganese oxides with a
  layered perovskite structure}}.
\newblock \emph{\bibinfo{journal}{Nature}} \textbf{\bibinfo{volume}{380}},
  \bibinfo{pages}{141--144} (\bibinfo{year}{1996}).

\bibitem{Maeno:1994cm}
\bibinfo{author}{Maeno, Y.} \emph{et~al.}
\newblock \bibinfo{title}{{Superconductivity in a layered perovskite without
  copper}}.
\newblock \emph{\bibinfo{journal}{Nature}} \textbf{\bibinfo{volume}{372}},
  \bibinfo{pages}{532--534} (\bibinfo{year}{1994}).

\bibitem{Mackenzie.75.657}
\bibinfo{author}{Mackenzie, A.~P.} \& \bibinfo{author}{Maeno, Y.}
\newblock \bibinfo{title}{{The superconductivity of Sr$_2$RuO$_4$ and the
  physics of spin-triplet pairing}}.
\newblock \emph{\bibinfo{journal}{Rev. Mod. Phys.}}
  \textbf{\bibinfo{volume}{75}}, \bibinfo{pages}{657--712}
  (\bibinfo{year}{2003}).

\bibitem{Callaghan1966}
\bibinfo{author}{Callaghan, A.}, \bibinfo{author}{Moeller, C.~W.} \&
  \bibinfo{author}{Ward, R.}
\newblock \bibinfo{title}{Magnetic interactions in ternary ruthenium oxides}.
\newblock \emph{\bibinfo{journal}{Inorg. Chem.}} \textbf{\bibinfo{volume}{5}},
  \bibinfo{pages}{1572--1576} (\bibinfo{year}{1966}).

\bibitem{shai2013}
\bibinfo{author}{Shai, D.~E.} \emph{et~al.}
\newblock \bibinfo{title}{{Doping and temperature dependence of the mass
  enhancement observed in the cuprate Bi$_2$Sr$_2$CaCu$_2$O$_{8+\delta}$}}.
\newblock \emph{\bibinfo{journal}{Phys. Rev. Lett.}}
  \textbf{\bibinfo{volume}{110}}, \bibinfo{pages}{087004--087008}
  (\bibinfo{year}{2013}).

\bibitem{Thiel:2006eo}
\bibinfo{author}{Thiel, S.}, \bibinfo{author}{Hammerl, G.},
  \bibinfo{author}{Schmehl, A.}, \bibinfo{author}{Schneider, C.~W.} \&
  \bibinfo{author}{Mannhart, J.}
\newblock \bibinfo{title}{{Tunable quasi-two-dimensional electron gases in
  oxide heterostructures}}.
\newblock \emph{\bibinfo{journal}{Science}} \textbf{\bibinfo{volume}{313}},
  \bibinfo{pages}{1942--1945} (\bibinfo{year}{2006}).

\bibitem{Cen:2008gr}
\bibinfo{author}{Cen, C.} \emph{et~al.}
\newblock \bibinfo{title}{{Nanoscale control of an interfacial metal--insulator
  transition at room temperature}}.
\newblock \emph{\bibinfo{journal}{Nature Mater.}} \textbf{\bibinfo{volume}{7}},
  \bibinfo{pages}{298--302} (\bibinfo{year}{2008}).

\bibitem{Huijben:ADFM201203355}
\bibinfo{author}{Huijben, M.} \emph{et~al.}
\newblock \bibinfo{title}{Defect engineering in oxide heterostructures by
  enhanced oxygen surface exchange}.
\newblock \emph{\bibinfo{journal}{Adv. Funct. Mater.}}
  \textbf{\bibinfo{volume}{23}}, \bibinfo{pages}{5240--5248}
  (\bibinfo{year}{2013}).

\bibitem{Brinkman:2007fk}
\bibinfo{author}{Brinkman, A.} \emph{et~al.}
\newblock \bibinfo{title}{{Magnetic effects at the interface between
  non-magnetic oxides}}.
\newblock \emph{\bibinfo{journal}{Nature Mater.}} \textbf{\bibinfo{volume}{6}},
  \bibinfo{pages}{493--496} (\bibinfo{year}{2007}).

\bibitem{Reyren31082007}
\bibinfo{author}{Reyren, N.} \emph{et~al.}
\newblock \bibinfo{title}{{Superconducting interfaces between insulating
  oxides}}.
\newblock \emph{\bibinfo{journal}{Science}} \textbf{\bibinfo{volume}{317}},
  \bibinfo{pages}{1196--1199} (\bibinfo{year}{2007}).

\bibitem{Kozuka:2009ht}
\bibinfo{author}{Kozuka, Y.} \emph{et~al.}
\newblock \bibinfo{title}{{Two-dimensional normal-state quantum oscillations in
  a superconducting heterostructure}}.
\newblock \emph{\bibinfo{journal}{Nature}} \textbf{\bibinfo{volume}{462}},
  \bibinfo{pages}{487--490} (\bibinfo{year}{2009}).

\bibitem{Tilley:1977gz}
\bibinfo{author}{Tilley, R.}
\newblock \bibinfo{title}{{Correlation between dielectric constant and defect
  structure of non-stoichiometric solids}}.
\newblock \emph{\bibinfo{journal}{Nature}} \textbf{\bibinfo{volume}{269}},
  \bibinfo{pages}{229--231} (\bibinfo{year}{1977}).

\bibitem{Udayakumar1988}
\bibinfo{author}{Udayakumar, K.~R.} \& \bibinfo{author}{Cormack, A.~N.}
\newblock \bibinfo{title}{{Structural aspects of phase equilibria in the
  strontium-titanium-oxygen system}}.
\newblock \emph{\bibinfo{journal}{J. Am. Ceram. Soc.}}
  \textbf{\bibinfo{volume}{71}}, \bibinfo{pages}{C469--C471}
  (\bibinfo{year}{1988}).

\bibitem{Udayakumar198955}
\bibinfo{author}{Udayakumar, K.} \& \bibinfo{author}{Cormack, A.}
\newblock \bibinfo{title}{{Non-stoichiometry in alkaline earth excess alkaline
  earth titanates }}.
\newblock \emph{\bibinfo{journal}{J. Phys. Chem. Solids}}
  \textbf{\bibinfo{volume}{50}}, \bibinfo{pages}{55--60}
  (\bibinfo{year}{1989}).

\bibitem{Dijkkamp1987}
\bibinfo{author}{Dijkkamp, D.} \emph{et~al.}
\newblock \bibinfo{title}{{Preparation of Y--Ba--Cu oxide superconductor thin
  films using pulsed laser evaporation from high T$_c$ bulk material}}.
\newblock \emph{\bibinfo{journal}{Appl. Phys. Lett.}}
  \textbf{\bibinfo{volume}{51}}, \bibinfo{pages}{619--621}
  (\bibinfo{year}{1987}).

\bibitem{Koinuma:1987jf}
\bibinfo{author}{Koinuma, H.} \emph{et~al.}
\newblock \bibinfo{title}{{Preparation of superconducting thin films of
  (La$_{1-x}$Sr$_{x}$)$_y$CuO$_{4-{\delta}}$ by sputtering}}.
\newblock \emph{\bibinfo{journal}{J. Appl. Phys.}}
  \textbf{\bibinfo{volume}{62}}, \bibinfo{pages}{1524--1526}
  (\bibinfo{year}{1987}).

\bibitem{Lathrop:1987hl}
\bibinfo{author}{Lathrop, D.~K.}, \bibinfo{author}{Russek, S.~E.} \&
  \bibinfo{author}{Buhrman, R.~A.}
\newblock \bibinfo{title}{{Production of YBa$_2$Cu$_3$O$_{7-y}$ superconducting
  thin films in situ by high-pressure reactive evaporation and rapid thermal
  annealing}}.
\newblock \emph{\bibinfo{journal}{Appl. Phys. Lett.}}
  \textbf{\bibinfo{volume}{51}}, \bibinfo{pages}{1554--1556}
  (\bibinfo{year}{1987}).

\bibitem{Poppe1988661}
\bibinfo{author}{Poppe, U.} \emph{et~al.}
\newblock \bibinfo{title}{{Direct production of crystalline superconducting
  thin films of YBa$_2$Cu$_3$O$_7$ by high-pressure oxygen sputtering}}.
\newblock \emph{\bibinfo{journal}{Solid State Commun.}}
  \textbf{\bibinfo{volume}{66}}, \bibinfo{pages}{661--665}
  (\bibinfo{year}{1988}).

\bibitem{Sandstrom:1988bm}
\bibinfo{author}{Sandstrom, R.~L.} \emph{et~al.}
\newblock \bibinfo{title}{{Reliable single-target sputtering process for
  high-temperature superconducting films and devices}}.
\newblock \emph{\bibinfo{journal}{Appl. Phys. Lett.}}
  \textbf{\bibinfo{volume}{53}}, \bibinfo{pages}{444--446}
  (\bibinfo{year}{1988}).

\bibitem{Spah:1988gi}
\bibinfo{author}{Spah, R.~J.}, \bibinfo{author}{Hess, H.~F.},
  \bibinfo{author}{Stormer, H.~L.}, \bibinfo{author}{White, A.~E.} \&
  \bibinfo{author}{Short, K.~T.}
\newblock \bibinfo{title}{{Parameters for in situ growth of high Tc
  superconducting thin films using an oxygen plasma source}}.
\newblock \emph{\bibinfo{journal}{Appl. Phys. Lett.}}
  \textbf{\bibinfo{volume}{53}}, \bibinfo{pages}{441--443}
  (\bibinfo{year}{1988}).

\bibitem{Madhavan:1996es}
\bibinfo{author}{Madhavan, S.}, \bibinfo{author}{Schlom, D.~G.},
  \bibinfo{author}{Dabkowski, A.}, \bibinfo{author}{Dabkowska, H.~A.} \&
  \bibinfo{author}{Liu, Y.}
\newblock \bibinfo{title}{{Growth of epitaxial a-axis and c-axis oriented
  Sr$_2$RuO$_4$ films}}.
\newblock \emph{\bibinfo{journal}{Appl. Phys. Lett.}}
  \textbf{\bibinfo{volume}{68}}, \bibinfo{pages}{559--561}
  (\bibinfo{year}{1996}).

\bibitem{Schlom:1999eo}
\bibinfo{author}{Schlom, D.~G.} \emph{et~al.}
\newblock \bibinfo{title}{{Growth of epitaxial films}}.
\newblock \emph{\bibinfo{journal}{Supercond. Sci. Technol.}}
  \textbf{\bibinfo{volume}{10}}, \bibinfo{pages}{891--895}
  (\bibinfo{year}{1999}).

\bibitem{Schlom1998}
\bibinfo{author}{Schlom, D.~G.} \emph{et~al.}
\newblock \bibinfo{title}{{Searching for superconductivity in epitaxial films
  of copper-free layered oxides with the K$_2$NiF$_4$ structure}}.
\newblock \emph{\bibinfo{journal}{Proc. SPIE}} \textbf{\bibinfo{volume}{3481}},
  \bibinfo{pages}{226--240} (\bibinfo{year}{1998}).

\bibitem{Matsuno:2003cu}
\bibinfo{author}{Matsuno, J.}, \bibinfo{author}{Okimoto, Y.},
  \bibinfo{author}{Kawasaki, M.} \& \bibinfo{author}{Tokura, Y.}
\newblock \bibinfo{title}{{Synthesis and electronic structure of epitaxially
  stabilized Sr$_{2-x}$La$_{x}$VO$_{4}$ $(0\leq x\leq1)$ thin films}}.
\newblock \emph{\bibinfo{journal}{Appl. Phys. Lett.}}
  \textbf{\bibinfo{volume}{82}}, \bibinfo{pages}{194--196}
  (\bibinfo{year}{2003}).

\bibitem{Shibuya:2008if}
\bibinfo{author}{Shibuya, K.}, \bibinfo{author}{Mi, S.}, \bibinfo{author}{Jia,
  C.-L.}, \bibinfo{author}{Meuffels, P.} \& \bibinfo{author}{Dittmann, R.}
\newblock \bibinfo{title}{{Sr$_2$TiO$_4$ layered perovskite thin films grown by
  pulsed laser deposition}}.
\newblock \emph{\bibinfo{journal}{Appl. Phys. Lett.}}
  \textbf{\bibinfo{volume}{92}}, \bibinfo{pages}{241918--241918}
  (\bibinfo{year}{2008}).

\bibitem{Tsuyoshi2011}
\bibinfo{author}{Ohnishi, T.} \& \bibinfo{author}{Takada, K.}
\newblock \bibinfo{title}{{Epitaxial thin-film growth of SrRuO$_3$,
  Sr$_3$Ru$_2$O$_7$, and Sr$_2$RuO$_4$ from a SrRuO$_3$ target by pulsed laser
  deposition}}.
\newblock \emph{\bibinfo{journal}{Appl. Phys. Express}}
  \textbf{\bibinfo{volume}{4}}, \bibinfo{pages}{025501} (\bibinfo{year}{2011}).

\bibitem{Feenstra:1995ei}
\bibinfo{author}{Feenstra, R.}, \bibinfo{author}{Budai, J.~D.},
  \bibinfo{author}{Christen, D.~K.} \& \bibinfo{author}{Kawai, T.}
\newblock \bibinfo{title}{{Superconductivity in
  Sr$_{n+1}$CunO$_{2n+1+{\delta}}$ ``infinite layer" films induced by
  postgrowth annealing}}.
\newblock \emph{\bibinfo{journal}{Appl. Phys. Lett.}}
  \textbf{\bibinfo{volume}{66}}, \bibinfo{pages}{2283--2285}
  (\bibinfo{year}{1995}).

\bibitem{Yoshiki1999}
\bibinfo{author}{Iwazaki, Y.}, \bibinfo{author}{Suzuki, T.},
  \bibinfo{author}{Sekiguchi, S.} \& \bibinfo{author}{Fujimoto, M.}
\newblock \bibinfo{title}{{Artificial SrTiO$_3$/SrO superlattices by pulsed
  laser deposition}}.
\newblock \emph{\bibinfo{journal}{Jpn. J. Appl. Phys.}}
  \textbf{\bibinfo{volume}{38}}, \bibinfo{pages}{L1443} (\bibinfo{year}{1999}).

\bibitem{Tanaka2000}
\bibinfo{author}{Tanaka, H.} \& \bibinfo{author}{Kawai, T.}
\newblock \bibinfo{title}{{Artificial construction of SrO/(La,Sr)MnO$_3$
  layered perovskite superlattice by laser molecular-beam epitaxy}}.
\newblock \emph{\bibinfo{journal}{Appl. Phys. Lett.}}
  \textbf{\bibinfo{volume}{76}}, \bibinfo{pages}{3618--3620}
  (\bibinfo{year}{2000}).

\bibitem{Schlom100443}
\bibinfo{author}{Schlom, D.~G.} \emph{et~al.}
\newblock \bibinfo{title}{{Molecular beam epitaxy of layered Dy--Ba--Cu--O
  compounds}}.
\newblock \emph{\bibinfo{journal}{Appl. Phys. Lett.}}
  \textbf{\bibinfo{volume}{53}}, \bibinfo{pages}{1660--1662}
  (\bibinfo{year}{1988}).

\bibitem{Bozovic:ez}
\bibinfo{author}{Bozovic, I.} \emph{et~al.}
\newblock \bibinfo{title}{{Atomic-layer engineering of cuprate
  superconductors}}.
\newblock \emph{\bibinfo{journal}{J. Supercond.}} \textbf{\bibinfo{volume}{7}},
  \bibinfo{pages}{187--195} (\bibinfo{year}{1994}).

\bibitem{LiuZY1994}
\bibinfo{author}{Liu, Z.}, \bibinfo{author}{Hanada, T.},
  \bibinfo{author}{Sekine, R.}, \bibinfo{author}{Kawai, M.} \&
  \bibinfo{author}{Koinuma, H.}
\newblock \bibinfo{title}{{Atomic layer control in Sr-Cu-O artificial lattice
  growth}}.
\newblock \emph{\bibinfo{journal}{Appl. Phys. Lett.}}
  \textbf{\bibinfo{volume}{65}}, \bibinfo{pages}{1717--1719}
  (\bibinfo{year}{1994}).

\bibitem{Haeni:2001iq}
\bibinfo{author}{Haeni, J.~H.} \emph{et~al.}
\newblock \bibinfo{title}{{Epitaxial growth of the first five members of the
  Sr$_{n+1}$Ti$_{n}$O$_{3n+1}$ Ruddlesden--Popper homologous series}}.
\newblock \emph{\bibinfo{journal}{Appl. Phys. Lett.}}
  \textbf{\bibinfo{volume}{78}}, \bibinfo{pages}{3292--3294}
  (\bibinfo{year}{2001}).

\bibitem{Okude2008}
\bibinfo{author}{Okude, M.}, \bibinfo{author}{Ohtomo, A.},
  \bibinfo{author}{Kita, T.} \& \bibinfo{author}{Kawasaki, M.}
\newblock \bibinfo{title}{{Epitaxial synthesis of Sr$_{n+1}$Ti$_n$O$_{3n+1}$
  ($n = 2-5$) Ruddlesden-Popper homologous series by pulsed-laser deposition}}.
\newblock \emph{\bibinfo{journal}{Appl. Phys. Express}}
  \textbf{\bibinfo{volume}{1}}, \bibinfo{pages}{081201} (\bibinfo{year}{2008}).

\bibitem{Tian:2007is}
\bibinfo{author}{Tian, W.} \emph{et~al.}
\newblock \bibinfo{title}{{Epitaxial growth and magnetic properties of the
  first five members of the layered Sr$_{n+1}$Ru$_n$O$_{3n+1}$ oxide series}}.
\newblock \emph{\bibinfo{journal}{Appl. Phys. Lett.}}
  \textbf{\bibinfo{volume}{90}}, \bibinfo{pages}{022507--022507}
  (\bibinfo{year}{2007}).

\bibitem{Yan:2007jl}
\bibinfo{author}{Yan, L.} \emph{et~al.}
\newblock \bibinfo{title}{{Unit-cell-level assembly of metastable
  transition-metal oxides by pulsed-laser deposition}}.
\newblock \emph{\bibinfo{journal}{Angew. Chem.}}
  \textbf{\bibinfo{volume}{119}}, \bibinfo{pages}{4623--4626}
  (\bibinfo{year}{2007}).

\bibitem{LeeCH1023}
\bibinfo{author}{Lee, C.-H.} \emph{et~al.}
\newblock \bibinfo{title}{{Effect of reduced dimensionality on the optical band
  gap of SrTiO$_3$}}.
\newblock \emph{\bibinfo{journal}{Appl. Phys. Lett.}}
  \textbf{\bibinfo{volume}{102}}, \bibinfo{pages}{122901}
  (\bibinfo{year}{2013}).

\bibitem{Lee:2013eg}
\bibinfo{author}{Lee, C.-H.} \emph{et~al.}
\newblock \bibinfo{title}{{Exploiting dimensionality and defect mitigation to
  create tunable microwave dielectrics}}.
\newblock \emph{\bibinfo{journal}{Nature}} \textbf{\bibinfo{volume}{502}},
  \bibinfo{pages}{532--536} (\bibinfo{year}{2013}).

\bibitem{Locquet1994}
\bibinfo{author}{Locquet, J.-P.}, \bibinfo{author}{Catana, A.},
  \bibinfo{author}{M{\"a}chler, E.}, \bibinfo{author}{Gerber, C.} \&
  \bibinfo{author}{Bednorz, J.~G.}
\newblock \bibinfo{title}{{Block-by-block deposition: A new growth method for
  complex oxide thin films}}.
\newblock \emph{\bibinfo{journal}{Applied physics letters}}
  \textbf{\bibinfo{volume}{64}}, \bibinfo{pages}{372--374}
  (\bibinfo{year}{1994}).

\bibitem{Haeni2000}
\bibinfo{author}{Haeni, J.~H.}, \bibinfo{author}{Theis, C.~D.} \&
  \bibinfo{author}{Schlom, D.~G.}
\newblock \bibinfo{title}{{RHEED intensity oscillations for the stoichiometric
  growth of SrTiO$_3$ thin films by reactive molecular beam epitaxy}}.
\newblock \emph{\bibinfo{journal}{J. Electroceram.}}
  \textbf{\bibinfo{volume}{4}}, \bibinfo{pages}{385--391}
  (\bibinfo{year}{2000}).

\bibitem{Tian:2011bc}
\bibinfo{author}{Tian, W.}, \bibinfo{author}{Pan, X.~Q.},
  \bibinfo{author}{Haeni, J.~H.} \& \bibinfo{author}{Schlom, D.~G.}
\newblock \bibinfo{title}{{Transmission electron microscopy study of $n= 1-5$
  Sr$_{n+1}$Ti$_n$O$_{3n+1}$ epitaxial thin films}}.
\newblock \emph{\bibinfo{journal}{J. Mater. Res.}}
  \textbf{\bibinfo{volume}{16}}, \bibinfo{pages}{2013--2026}
  (\bibinfo{year}{2011}).

\bibitem{Kleibeuker:ADFM201000889}
\bibinfo{author}{Kleibeuker, J.~E.} \emph{et~al.}
\newblock \bibinfo{title}{{Atomically defined rare-earth scandate crystal
  surfaces}}.
\newblock \emph{\bibinfo{journal}{Adv. Funct. Mater.}}
  \textbf{\bibinfo{volume}{20}}, \bibinfo{pages}{3490--3496}
  (\bibinfo{year}{2010}).

\bibitem{Koster122630}
\bibinfo{author}{Koster, G.}, \bibinfo{author}{Kropman, B.~L.},
  \bibinfo{author}{Rijnders, G. J. H.~M.}, \bibinfo{author}{Blank, D. H.~A.} \&
  \bibinfo{author}{Rogalla, H.}
\newblock \bibinfo{title}{{Quasi-ideal strontium titanate crystal surfaces
  through formation of strontium hydroxide}}.
\newblock \emph{\bibinfo{journal}{Appl. Phys. Lett.}}
  \textbf{\bibinfo{volume}{73}}, \bibinfo{pages}{2920--2922}
  (\bibinfo{year}{1998}).

\bibitem{shirley1972high}
\bibinfo{author}{Shirley, D.~A.}
\newblock \bibinfo{title}{High-resolution x-ray photoemission spectrum of the
  valence bands of gold}.
\newblock \emph{\bibinfo{journal}{Phys. Rev. B}} \textbf{\bibinfo{volume}{5}},
  \bibinfo{pages}{4709} (\bibinfo{year}{1972}).

\bibitem{Yu200451}
\bibinfo{author}{Yu, Z.} \emph{et~al.}
\newblock \bibinfo{title}{Advances in heteroepitaxy of oxides on silicon}.
\newblock \emph{\bibinfo{journal}{Thin Solid Films}}
  \textbf{\bibinfo{volume}{462â463}}, \bibinfo{pages}{51--56}
  (\bibinfo{year}{2004}).

\bibitem{Fisher2008JAP}
\bibinfo{author}{Fisher, P.} \emph{et~al.}
\newblock \bibinfo{title}{{Stoichiometric, nonstoichiometric, and locally
  nonstoichiometric SrTiO$_3$ films grown by molecular beam epitaxy}}.
\newblock \emph{\bibinfo{journal}{J. Appl. Phys.}}
  \textbf{\bibinfo{volume}{103}}, \bibinfo{pages}{013519}
  (\bibinfo{year}{2008}).

\bibitem{Lee2013APL}
\bibinfo{author}{Lee, C.-H.} \emph{et~al.}
\newblock \bibinfo{title}{{Effect of stoichiometry on the dielectric properties
  and soft mode behavior of strained epitaxial SrTiO$_3$ thin films on
  DyScO$_3$ substrates}}.
\newblock \emph{\bibinfo{journal}{Appl. Phys. Lett.}}
  \textbf{\bibinfo{volume}{102}}, \bibinfo{pages}{082905}
  (\bibinfo{year}{2013}).

\bibitem{Venkateswaran19871273}
\bibinfo{author}{Venkateswaran, U.}, \bibinfo{author}{Strössner, K.},
  \bibinfo{author}{Syassen, K.}, \bibinfo{author}{Burns, G.} \&
  \bibinfo{author}{Shafer, M.}
\newblock \bibinfo{title}{{Pressure dependence of the Raman modes in
  Sr$_2$TiO$_4$}}.
\newblock \emph{\bibinfo{journal}{Solid State Commun.}}
  \textbf{\bibinfo{volume}{64}}, \bibinfo{pages}{1273--1277}
  (\bibinfo{year}{1987}).

\bibitem{Loviat2007}
\bibinfo{author}{Loviat, F.} \emph{et~al.}
\newblock \bibinfo{title}{The surface layer of cleaved bilayer manganites}.
\newblock \emph{\bibinfo{journal}{Nanotechnology}}
  \textbf{\bibinfo{volume}{18}}, \bibinfo{pages}{044020}
  (\bibinfo{year}{2007}).

\bibitem{Matzdorf:2000jm}
\bibinfo{author}{Matzdorf, R.} \emph{et~al.}
\newblock \bibinfo{title}{{Ferromagnetism stabilized by lattice distortion at
  the surface of the p-wave superconductor Sr$_2$RuO$_4$}}.
\newblock \emph{\bibinfo{journal}{Science}} \textbf{\bibinfo{volume}{289}},
  \bibinfo{pages}{746--748} (\bibinfo{year}{2000}).

\bibitem{Junquera-03.2}
\bibinfo{author}{Junquera, J.}, \bibinfo{author}{Zimmer, M.},
  \bibinfo{author}{Ordej\'on, P.} \& \bibinfo{author}{Ghosez, P.}
\newblock \bibinfo{title}{{First-principles calculation of the band offset at
  BaO/BaTiO$_3$ and SrO/SrTiO$_3$}}.
\newblock \emph{\bibinfo{journal}{Phys. Rev. B}} \textbf{\bibinfo{volume}{67}},
  \bibinfo{pages}{155327} (\bibinfo{year}{2003}).

\bibitem{Theis:1996gz}
\bibinfo{author}{Theis, C.~D.} \& \bibinfo{author}{Schlom, D.~G.}
\newblock \bibinfo{title}{{Cheap and stable titanium source for use in oxide
  molecular beam epitaxy systems}}.
\newblock \emph{\bibinfo{journal}{J Vac. Sci. Technol. A}}
  \textbf{\bibinfo{volume}{14}}, \bibinfo{pages}{2677--2679}
  (\bibinfo{year}{1996}).

\bibitem{Voyles2003251}
\bibinfo{author}{Voyles, P.}, \bibinfo{author}{Grazul, J.} \&
  \bibinfo{author}{Muller, D.}
\newblock \bibinfo{title}{{Imaging individual atoms inside crystals with
  ADF-STEM}}.
\newblock \emph{\bibinfo{journal}{Ultramicroscopy}}
  \textbf{\bibinfo{volume}{96}}, \bibinfo{pages}{251--273}
  (\bibinfo{year}{2003}).

\bibitem{Kirkland20111523}
\bibinfo{author}{Kirkland, E.~J.}
\newblock \bibinfo{title}{{On the optimum probe in aberration corrected
  ADF-STEM}}.
\newblock \emph{\bibinfo{journal}{Ultramicroscopy}}
  \textbf{\bibinfo{volume}{111}}, \bibinfo{pages}{1523--1530}
  (\bibinfo{year}{2011}).

\bibitem{Livesey1994439}
\bibinfo{author}{Livesey, A.} \& \bibinfo{author}{Smith, G.}
\newblock \bibinfo{title}{{The determination of depth profiles from
  angle-dependent XPS using maximum entropy data analysis}}.
\newblock \emph{\bibinfo{journal}{J. Electron Spectrosc.}}
  \textbf{\bibinfo{volume}{67}}, \bibinfo{pages}{439--461}
  (\bibinfo{year}{1994}).

\bibitem{EAL}
\bibinfo{note}{C. J. Powell and A. Jablonski, NIST Electron
  Effective-Attenuation-Length Database - Version 1.1 (National Institute of
  Standards and Technology, Gaithersberg, MD, 2001).}

\bibitem{siesta}
\bibinfo{author}{Soler, J.~M.} \emph{et~al.}
\newblock \bibinfo{title}{{The Siesta method for {\it ab initio} order-{\it N}
  materials simulations}}.
\newblock \emph{\bibinfo{journal}{J. Phys.: Condens. Matter}}
  \textbf{\bibinfo{volume}{14}}, \bibinfo{pages}{2745--2779}
  (\bibinfo{year}{2002}).

\end{thebibliography}
\end{document}